\documentstyle[12pt]{article}    
\newcommand{\ctg}{\mathop{\rm ctg}\nolimits}

\begin{document} 
 
\begin{titlepage}    
    
\begin{flushright}    
Moscow University\\
Physics Bulletin\\
Vol. 51, No. 1. (1996) 1-6
\end{flushright}    
\bigskip    
    
\begin{center}    
\large\bf    
Radiative Shift of the Quark Mass in a Constant \\
Chromomagnetic Field at Finite\\
Temperature and Density
\end{center}    
\vspace{0.5cm}    
\begin{center}    
V.Ch. Zhukovskii, K. G. Levtchenko, T. L. Shoniya \\ 
{\sl  
Faculty of Physics,\\ 
Department of Theoretical Physics, Moscow State University,\\  
119899, Moscow, Russia} \\   
P. A. Eminov\\ 
{\sl  
Department of Physics\\ 
Moscow State Institute of Electronics and Mathematics\\ 
(Technical University)\\ 
109028, Moscow, Russia} 
\end{center}

\begin{abstract}    
The radiation shift of the quark mass in a constant chromomagnetic 
field at finite temperature and density was calculated. The limiting cases 
of a weak and a strong chromomagnetic field were considered. It was 
shown that in a strong field there is no imaginary part in the 
contribution of the finite density effects to the quark mass shift, and its 
real part can considerably exceed the corresponding part of the purely 
field contribution.
\end{abstract}    
    
\vspace{0.5cm}    
   
\vfill    
\end{titlepage}

In spite of considerable efforts that have been undertaken in recent 
years, there is no consistent theory of the QCD vacuum state at present. 
All recently proposed QCD vacuum models are based upon 
investigation of the interaction of quantized quark and gluon fields 
against the background of a gluon condensate, i.e., a classical gauge 
field. Along with instanton models of the QCD vacuum [1], efforts have 
been undertaken to construct models that are based on replacement of 
stochastic vacuum gluon fields by various regular configurations of 
external non--Abelian fields, the simplest of which is a constant 
chromomagnetic field of the Abelian type proposed by Matinyan and 
Savvidy~[2].

In this connection various radiation effects in non--Abelian external 
fields became the subject of intensive studies [3--5]. On the other hand, 
many problems of elementary particle physics have statistical aspects, 
such as problems in phase transitions in gauge theories at finite 
temperature and non--zero chemical potential [6], the question on the 
possibility of stabilizing the classical Yang--Mills field configurations 
through temperature effects [7], the problem of the particle energy 
radiative shift and modifications of the particle electromagnetic 
properties at the expense of finite temperature and density effects [8, 9], 
etc.

In [10], spectra of gluon and quark--like excitations in the hot 
quark--gluon plasma were calculated in the one--loop approximation, 
and it was pointed out that the results obtained could be used in the 
analysis of experimental data on collisions of relativistic nuclei (which 
are expected to generate the quark--gluon plasma [11]) and also in the 
analysis of the processes that take place at the early stages of evolution 
of the Universe and in quark stars.

In the present paper we consider the contribution of the finite density 
effects to the radiation shift of the quark energy with consideration for 
the vacuum  fluctuations of gauge bosons. The vacuum state here is 
approximated by the Matinyan and Savvidy constant chromomagnetic 
field.

In the real time representation used in the present paper, the quark 
energy radiative shift is written in the form of the sum of two terms, one 
of which corresponds to the energy radiative shift at $T=\mu=0$, and 
the other term corresponds to the contributions of the finite density and 
temperature effects~[12]. 

Following [3], we represent the total gluon field $A^a_{\mu}$ as a sum 
of a classical (non--Abelian) external field $\bar A^a_{\mu}$ and small 
quantum fluctuations around it: 
$$
  A^a_{\mu}=\bar A^a_{\mu}+Q^a_{\mu}.
  $$
The external field is chosen in the Matinyan--Savvidy form specified by 
the potential
\begin{equation}
  \bar A^a_{\mu}=\delta^{a8}A_{\mu},\quad 
A_{\mu}=B\delta^2_{\mu}x_1,
  \label{aa}
  \end{equation}
and the interaction of the quark with the external field $\bar 
A^a_{\mu}$ will be taken into account exactly, while its interaction with 
the quantized gluon field $Q^a_{\mu}$ will be treated by the 
perturbation theory. 

The one--loop energy shift of the color $k$ quark in the external 
chro\-mo\-mag\-ne\-tic field has the form~[3--5] 
\begin{equation}
  \triangle E=\frac1T \int d^4xd^4x'\bar\psi_k(x)M_{(k)}(x,x')\psi_k(x'),
  \label{ve}
  \end{equation}
where the mass operator, diagonal with respect to color in view of the 
specific choice of the external field (1), is defined by the formulas
$$
  M_{kl}(x,x')=\delta_{kl}M_{(k)}(x,x')
  $$
  \begin{equation}
  M_{(k)}(x,x')=-ig^2 \gamma^{\mu}{\left({\hat T}_a 
\right)}_{kn}S_{(n)}(x,x')
         \gamma^{\nu}{\left({\hat T}^{+}_a 
\right)}_{nk}D^{(a)}_{\mu\nu}(x,x').
  \label{mas}
  \end{equation}

In the following, for definiteness,  we will consider, as in [5], the case 
when the quark color is $k=1$, i.e., its charge is $e_1=\bar g/3$. Then 
>from Eqs. (2), (3) and the explicit form of matrices $T_a=\lambda_a/2$ 
it follows that both neutral $(a=1, 2, 3, 8)$ and charged gluons 
contribute to the quark energy shift. In the most interesting case of a 
charged gluons contribute to the quark energy shift. In the most 
interesting case of a charged gluon  $(a=4,\quad g_a=\bar g)$, due to 
the law of conservation of the color charge, the intermediate quark has 
the charge $e_n=-2\bar g/3\quad(n=3)$.  This is just the case we 
consider here. 

In the real time representation, in order to compute the one--loop 
con\-tribu\-tions of the finite density and temperature to the quark energy 
shift one has to replace in formula (3) the causal Green functions 
$S_{(n)}(x,x')$ and $D^{(a)}_{\mu\nu}(x,x')$ by the time Green 
functions of the ideal quark--antiquark and gluon gases~[13].

The following representation for the time Green function [8] will be used: 
\begin{equation}
  S(x,x')=S_{(n)}(x,x')
  -\sum_{\{s\},\varepsilon=\pm1}\frac{
  \varepsilon\Psi^{(\varepsilon)}_s(\vec x)
  {\bar\Psi}^{(\varepsilon)}_s({\vec x}^{\,\prime})}
  {\exp{\left\{\frac{E_s-\varepsilon\mu}{T}\right\}}+1}
  \exp{\{-i\varepsilon E_n(x_0-x^\prime_0)\}}.
  \label{qpn}
  \end{equation}
Here $\Psi^{(\pm)}_s$ and $E^{(\pm)}_s$ are, respectively, the wave 
functions and energy levels in the constant chromomagnetic field (1) of 
the quark and antiquark, summation is performed over all quantum 
states ${s}$ of quarks $(\varepsilon=1)$ and antiquarks $(\varepsilon=-
1)$, and the first term is the causal Green function of a quark at 
$T=\mu=0$~[3--5].

Thus, the calculation of the quark energy shift in the case of a 
degenerate quark gas is reduced to the replacement in (3) of the causal 
Green function by the time Green function (4), and the explicit form of 
the causal Green function of the charged gluon in (3) is presented in~[5].

As a result, the ground state energy shift of a quark of the color $k=1$ 
is represented by the sum of two terms: 
\begin{equation}
  \triangle E=\triangle E(H,T=\mu=0)+\triangle E(H,\mu\ne0,T\ne0).
  \label{tel}
  \end{equation}
The first term in (5) corresponds to the radiative energy shift of the 
quark in the field (1) at  $T=\mu=0$ and it has been considered in 
detail in [3--5]; the second term describes the contribution of the 
exchange interaction to the quark energy shift and is of  interest to us.

Calculating the quantity $\triangle E(H,T\ne0,\mu\ne0)$ leads to the 
expression
\begin{equation}
  \triangle E(H,T,\mu)=-g^2m\frac{H_0}H\frac 
i{(2\pi)^3}\sum_{\varepsilon}
  \int\limits^{\infty}_0r^2\,dr
  \int\limits^{\infty}_0\,dt\int\limits^{+\infty}_{-
\infty}ds\varepsilon\,A(r,s,t).
  \label{det}
  \end{equation}
Here
$$
A(r,s,t)=\frac{\sqrt{\pi}}2\frac{\exp{\{i(s+t)\rho^2\}}}{\rho\sqrt{i(s+t)}}
  \Phi(\rho\sqrt{i(s+t)})
  \frac1{\exp{\left\{\frac{\sqrt{m^2+\rho^2}}{T}\right\}}+1}\times
  $$
  $$
  \times\frac{\exp{\left\{2it[m^2-\varepsilon 
m\sqrt{m^2+\rho^2}]\right\}}}{\sin\bar gHt
  \sin\frac23\bar gHs}\;
  \frac1{\frac23\bar gH\ctg\frac23\bar gHs+\bar gH\ctg\bar gHt+
  \frac{i\bar gH}3}\times
  $$
  $$
  \times\left[\frac{m\exp{\left\{-\frac23i\bar 
gHs\right\}}}{\sqrt{m^2+\rho^2}}-
  \left(1-\frac{m}{\sqrt{m^2+\rho^2}}\right)\exp{\left\{\frac23i\bar 
gHs\right\}}
  \exp{\{2i\bar gHt\}}\right],
  $$
where $\Phi(z)$ is the error integral and $\rho=m r$.

The case of fermions with the isotopic spin 1  in the theory with the 
$SU(2)$ gauge group (the adjoint representation) is considered in a 
similar way. It is well known [3, 4] that in the representation chosen, 
fermions and bosons have the following charges with respect to the 
external field: $-g,0,g$. We will restrict ourselves to the case of fermions 
with the charge $g$  in the ground state which, as well as the case 
considered above in the framework of QCD, does not have any 
electrodynamical analog. Then the intermediate boson in the one--loop 
diagram for the radiation correction to the energy has the charge $+g$, 
and the intermediate fermion is neutral.

The contribution of the exchange interaction to the fermion energy shift 
is calculated according to the above scheme:
\begin{equation}
  \triangle E(H,T,\mu)=-
g^2m\frac{H_0}H\frac{2i}{(2\pi)^3}\sum_{\varepsilon}
  \int\limits^{\infty}_0r^2\,dr\int\limits^{\infty}_0d\tau\varepsilon 
A(r,\tau),
  \label{etw}
  \end{equation}
where
  $$
  A(r,\tau)=\frac{\sqrt{\pi}}2\frac{\Phi(Z)}Z\exp{\{Z^2\}}
  \exp{\left\{-2i\tau\frac{H_0}H[\varepsilon\sqrt{1+r^2}-1]\right\}}\times
  $$
  $$
  \times\frac1{\exp{\left\{\frac{m(\sqrt{r^2+1}-
\frac{\mu}m)}T\right\}}+1}
  \left[\frac{\exp{\{-i\tau\}}}{\sqrt{r^2+1}}-\left(1-
\frac1{\sqrt{r^2+1}}\right)
  \exp{\{i\tau\}}\right]
  $$
  \smallskip
and the following notation is adopted
  $$
  H_0=\frac{m^2}{\bar g},\qquad
  \tau=\bar g Ht,\qquad
  Z=-ir^2\frac{H_0}H(\sin\tau\exp{\{-i\tau\}}-\tau).
  $$
 
In the limiting case of a highly degenerate quark gas, when
$$
T\ll E_F=\mu\quad(T=0),
$$
the first term in the expansion in formula (5) with respect to the 
parameter $T/E_F\ll1$ corresponds to the replacement of the Fermi 
distribution by the $\theta$--function~[14]:
$$
\frac1{\exp{\left\{\frac{m(\sqrt{r^2+1}-\frac{\mu}m)}T\right\}}+1}
  \longrightarrow\theta\left(\frac{\mu}m-\sqrt{r^2+1}\right)
  =\left\{
  \begin{array}{l}
  0,\;\mu<m\sqrt{r^2+1}\\
  1,\;\mu>m\sqrt{r^2+1}
  \end{array}\right.
  $$

In the limiting case of zero temperature which is of interest to us, when 
there  are no antiqurks in the gas, only the contribution of the positive--
frequency states should be left in formulas (6) and (7).

We consider first the energy shift of a quark in the QCD in a 
comparatively strong chromomagnetic field, when the following 
condition
\begin{equation}
  2|e_3|H>\mu^2-m^2
  \label{str}
  \end{equation}
is fulfilled. For the contribution of the finite density effects to the energy 
shift of the quark ground state we obtain the following exact results:
\begin{equation}
  \triangle E=m\frac{g^2}{6\pi^2}\left(\frac{\bar gH}{m^2}\right)
  \left\{\ln\left(y+\sqrt{y^2-1}\right)+
  x\frac1{\displaystyle\sqrt{x(2+x)}}\times\right.
  \label{ebh}
  \end{equation}
$$
  \left.\times\ln\left|\frac{1-y(1+x)+\sqrt{(y^2-1)x(2+x)}}{y-1-x}
  \right|_{y=\frac{\mu}m}\right\},
  \quad x=\frac{\displaystyle\bar gH}{\displaystyle 2m^2}.
$$
  
We note that the chemical potential, or, to be exact, the Fermi energy in 
formula (9) are related to the quark concentration by the relationship
\begin{equation}
  n_e=\frac{|e_3|H}{(2\pi)^2}
  \int\limits^{+\sqrt{\mu^2-m^2}}_{-\sqrt{\mu^2-m^2}}dp_3=
  \frac{|e_3|H}{2\pi^2}\sqrt{\mu^2-m^2}.
  \label{conqa}
  \end{equation}
In the limiting case of a strong field when
$$
\sqrt{\left(\frac{\mu}m\right)^2-1}\ll\frac{H}{H_0},
$$
it follows from Eq. (9) with regard for (10) that
$$
\triangle m=-m\frac{g^2}{2\pi^2}\sqrt{\frac{H_0}H}
\left(\frac{n_e}{m^3}\right)^{1/2}.
$$
Now we pass to the case of a comparatively weak chromomagnetic 
field:
\begin{equation}
  H\ll H_0,\quad 2|e_3|H\ll\mu^2-m^2.
  \label{rwf}
  \end{equation}
In the free case when $H=0$ we find from (6)
\begin{equation}
  \triangle m(H=0,T=0,\mu\ne0)=\frac{\alpha_s}{4\pi}m
  \left\{\sqrt{\left(\frac{\mu}m\right)^2-1}\left(\frac{\mu}m-2\right)-\right.
  \label{den}
  \end{equation}
$$
  \left.-3\ln\left(\frac{\mu}m+\sqrt{\left(\frac{\mu}m\right)^2-
1}\;\right)\right\},
$$
where $\alpha_s=g^2/(4\pi)$ and the chemical potential of the free 
quark gas is related to the density $n$ by the equation~[14]
$$
\frac{\mu}m=\left[1+\left(\frac{3\pi^2n}{m^3}\right)^{2/3}
\right]^{1/2}.
$$

After the replacement $\alpha_s/2\to\alpha_s$, the result (12) coincides 
with the corresponding result for the theory with the $SU(2)$ gauge 
group, and the latter, as was to be expected after the replacement 
$\alpha_s\to\alpha$ ($\alpha$ is the fine structure constant), coincides 
with the well--known expression for the contribution of finite density 
effects to the electron mass shift in the QED~[8].

To calculate the field contribution to the quark mass shift in the case of 
the $SU(2)$ gauge theory, when conditions (11) are fulfilled, one has to 
make use of the known representation of the error integral in the form of 
the series~[15]
\begin{equation}
  \Phi(z)=\frac2{\sqrt{\pi}}\exp{\{-
z^2\}}\sum^{\infty}_{n=0}\frac{2^nz^{2n+1}}{(2n+1)!!}.
  \label{int}
  \end{equation}

When conditions (11) are fulfilled, the first term in series (13) 
corresponds to the main term in the asymptotics (6). In particular, in the 
limiting case of low densities, when
\begin{equation}
  \frac{H}{H_0}\ll\left(\frac{\mu}m\right)^2-1\ll 1,
  \label{wn}
  \end{equation}
we find that
  \begin{equation}
  \triangle m(H\ne0,\mu\ne0,T=0)=-2m\frac{\alpha_s}{\pi}
  \sqrt{\left(\frac{\mu}m\right)^2-1}\left[1-
  \sqrt{\frac{gH}{\mu^2-m^2}}\right].
  \label{wecon}
  \end{equation}

As seen from (15), the contribution of the finite density effects to the 
quark mass shift is negative in the case of low densities, whereas the 
purely field contribution is positive. In the other limiting case, when 
$\mu/m\gg1$, we have
$$
\triangle m=\frac{\alpha_s}{4\pi}m\left(\frac{\mu}m\right)^2
  \left[1+\frac{2gH}{m\mu}\right].
  \label{ln}
$$

Thus, as the density of the quark gas grows, the radiative shift of the 
quark mass, which is negative at low densities, passes the zero value and 
increases proportionally to $\left(\frac{\mu}m\right)^2$ at high densities, 
whereas the leading field contribution is positive both at low and high 
densities.

We also note that up to the replacement 
$\alpha_s\longrightarrow\alpha_s/2$ the asymptotics (15) is also valid in 
the case of the $SU(3)$ gauge group. Similar to the calculation of the 
radiative shift of the quark mass at $T=\mu=0$ 
[3--5],  this is related to the fact that a weak external field 
$H\ll H_0=m^2/g$ turns out to be weak for the quark but superstrong 
for the massless charged gluon.

Comparison of the results obtained with the radiative correction to the 
quark mass shift in a constant chromomagnetic field at zero temperature 
and nonzero chemical potential [3--5] shows that under the condition 
(14) the field correction, as in the case of   $T=\mu=0$ has, according to 
(15), the order of magnitude $(gH)^{\frac12}$ though the sign is 
opposite.

As was first reported in [3], the ground state of a quark in the 
cromo\-mag\-ne\-tic field (1) at  $T=\mu=0$ has a positive imaginary part, 
i.e., it is unstable. This instability is caused by the presence of a 
tachionic mode in the gluon energy spectrum in such a field.

In [3], the imaginary part of the quantity $\triangle E_0$ is interpreted in 
the following way: as a result of gluon absorption from the gluon pair 
created by the unstable external field, the quark passes from the ground 
to an excited state. It seems interesting that in the case of a 
comparatively strong magnetic field (when condition (8) is fulfilled) no 
imaginary part in the energy shift of the ground state of a quark arises 
on account of the finite density effects. This is immediately seen from the 
exact result (9) and it can be explained by the fact that all quarks under 
condition (8) may stay only in the ground state, so that the
above--mentioned transitions to excited states on account of gluon absorption 
cannot take place. At the same time in a comparatively weak magnetic 
field, when conditions (11) are fulfilled, allowance for the terms of series 
(13) that follow the main term leads to the appearance of an imaginary 
part in the contribution of the exchange interaction to the quark energy 
shift, too. This can be qualitatively explained by the fact that transitions 
of the quark to excited states are no longer forbidden under condition (11).

The authors are grateful to A.V. Borisov for the discussion of the results 
of this work.

\end{document}